\documentclass[preprint,showpacs,preprintnumbers,amsmath,amssymb]{revtex4}
\usepackage{graphicx}

\begin{document}

\title{Majorana Neutrinos Production at NLC in an Effective Approach}

\author{Javier Peressutti}
\affiliation{Instituto de F\'{\i}sica de Mar del Plata (IFIMAR)\\
CONICET, UNMDP\\ Departamento de F\'{\i}sica,
Universidad Nacional de Mar del Plata \\
Funes 3350, (7600) Mar del Plata, Argentina}

\author{Ismael Romero}

\author{Oscar A. Sampayo}
\email{sampayo@mdp.edu.ar}

 \affiliation{Instituto de F\'{\i}sica de Mar del Plata (IFIMAR)\\ CONICET, UNMDP\\ Departamento de F\'{\i}sica,
Universidad Nacional de Mar del Plata \\
Funes 3350, (7600) Mar del Plata, Argentina}

\begin{abstract}
We investigate the possibility of detecting Majorana neutrinos at
the $e^+e^-$ Next Linear Collider (NLC).
 We study the $l_j^{\mp} l_k^{\mp} +  jets$ ($l_j\equiv e ,\mu
,\tau$)
 final states which are, due to leptonic
 number violation, a clear signature for intermediate Majorana
 neutrino contributions. Such signals (final leptons of the same-sign) are not
 possible if the heavy
 neutrinos have Dirac nature.  The interactions between Majorana neutrinos
 and the Standard Model (SM) particles are obtained from an effective
 Lagrangian approach. As for the background, we considered the SM reaction
 $e^+ e^- \rightarrow W^+W^+W^-W^-$, with two $W^\prime$s decaying into jets and two $W^\prime$s decaying into
 $l^{\pm}+\nu(\bar \nu)$, producing extra light neutrinos which avoid
 the detection.
 We present our results for the total
 cross-section as a function of the neutrino mass and the center of mass
 energies. We also show the discovery region as a function of the Majorana
 neutrino mass and the effective coupling.
\end{abstract}

 \maketitle
\maketitle

\section{\bf Introduction}
The standard model of particle physics (SM) only contains
left-handed neutrinos, which makes it not possible to generate mass
for them. One very important discovery in the field is the neutrino
oscillations, which requires the neutrinos to posses a small mass
($m_{\nu}\gtrsim 0.01$ eV). Thus we need to go beyond the SM to
solve this issue. One way to do it, is by the seesaw mechanism,
which requires one o more right-handed neutrinos, generically
$\nu_R$, with a mass term
\begin{eqnarray}
\label{lmass} \mathcal L^{mass}= - \frac12 \bar{\nu_R^c} \;  M \;
\nu_R - \bar L \; \widetilde{\phi} \; Y  \; \nu_R + h.c. \; ,
\end{eqnarray}
where $L$ denotes the left-handed lepton doublet, $Y$ the Yukawa
coupling matrix , $\phi$ the doublet higgs boson and $M$ the
Majorana mass matrix. There are many extensions of the SM
(Left-right symmetric model, SO(10), E6, ...) with extra
right-handed neutrinos which are singlets of the SM gauge group and
for which the Majorana mass terms are naturally allowed
\cite{neutmass}.

Upon diagonalization of the mass term we obtain, besides the light
neutrinos, heavy Majorana neutrinos (N), which allow for lepton
number violation.

By solving the eigenvalue problem, we obtain the masses
\begin{eqnarray}
\label{mnu} m_{\nu}=m_{\mathcal D} M^{-1} m^T_{\mathcal D}, \;\;\;
\mbox{with} \;\;\; m_{\mathcal D} = Y \frac{v}{\sqrt{2}} \; ,
\end{eqnarray}
and the mixing angle $U_{lN} \sim m_{\mathcal D}/M$. In typical
seesaw scenarios, the Dirac mass term $m_{\mathcal D}$ are expected
to be around the electroweak scale (then $Y\sim O(1)$ in
Eq.(\ref{mnu})), whilst the Majorana mass $M$ being a singlet under
the SM gauge group may be very large, close to the Grand Unification
Scale. Thus, the seesaw mechanism can explain the smallness of the
observed light neutrino masses ($m_{\nu}\sim0.01$ eV) and clearly
leads to the decoupling of $N$. Even a different choice in which
$M\sim100$ GeV and $m_{\mathcal D}\sim 0.1\; m_{e}$, keeping
$m_{\nu}\sim 0.01$ eV, implies a vanishing mixing $U_{lN}\sim
10^{-7}$ \cite{Aguila1}.

This mixing weighs the coupling of $N$ with the standard model
particles and in particular with the charged leptons through the
$V-A$ interaction:
\begin{eqnarray}
\label{lw}
 \mathcal L_W = -\frac{g}{\sqrt{2}} U_{lN} \bar N^c
\gamma^{\mu} P_L l W^+_{\mu} + h.c
\end{eqnarray}
This effect is so weak that the observation of Lepton Number
Violation (LNV) must indicate new physics beyond the minimal seesaw
mechanism, as was indicated in Ref.\cite{Aguila1}. In view of the
above discussion we consider, in a model independent way, the
effective interactions of the Majorana neutrino $N$, of mass lower
than $1$ TeV and negligible mixing to $\nu_L$ .

In the case that heavy neutrinos ($N$) exist, the present and future
experiments will be capable of determining their nature. The
production of Majorana neutrinos via $e^+e^-$, $e^- \gamma$, $\gamma
\gamma$ and hadronic collision have been extensively investigated in
the past \cite{Aguila1,ma,datta,gluza,hoefer,cevetic,almeida,
jperessu,jperessu1,belanger,atre}.

\begin{figure*}[htbp]
\begin{center}
\includegraphics[totalheight=5cm]{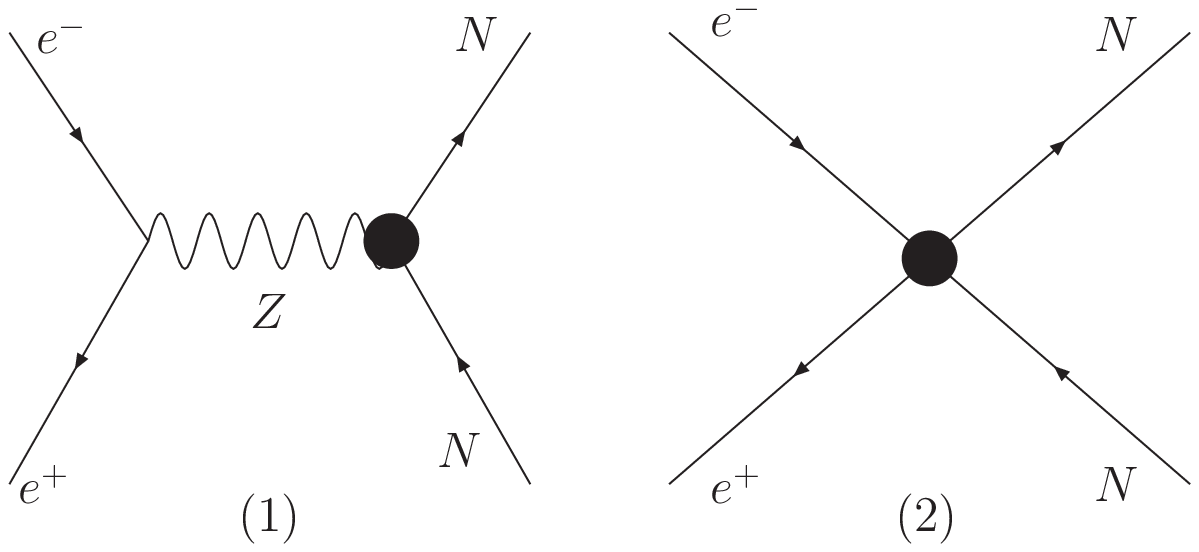}
\caption{\label{fig:eeNN}  Diagram contributing to the production of
two Majorana neutrinos.}
\end{center}
\end{figure*}

\begin{figure*}[htbp]
\begin{center}
\includegraphics[totalheight=5cm]{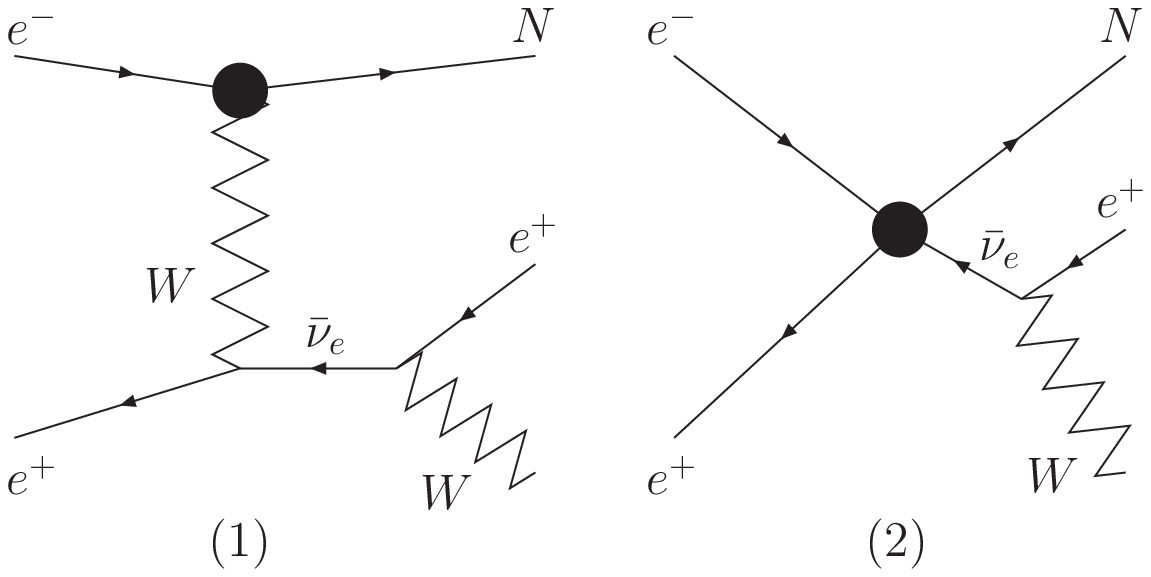}
\caption{\label{fig:eeN}  Diagram contributing to the production of
a single Majorana neutrino.}
\end{center}
\end{figure*}

In this work we study the possibility for the $e^+e^-$ next linear
collider (NLC) to produce clear signatures of Majorana neutrinos in
the context of interactions coming from an effective lagrangian
approach. We study the reaction $e^+ e^- \rightarrow l_j^{\mp}
l_k^{\mp} + jets$ ($l_j\equiv e,\mu,\tau$), which is divided into
two subprocesses depicted in Fig.\ref{fig:eeNN} and
Fig.\ref{fig:eeN}. In the first case we produce two Majorana
neutrinos ($N$) which will decay into one charged lepton and jets
($N\rightarrow l+ jets$). In the second case, which is a three body
reaction, we consider single neutrino production decaying in the
same way as before, and a $W$ decaying into two jets ($W \rightarrow
jets$). We have not considered the pure lepton channels since they
involve light neutrinos which escape detection, in which case the
Majorana nature of the heavy neutrinos would have no effect on the
signal (we should be able to know whether the final state contains
neutrinos or antineutrinos). For the decay of the Majorana neutrinos
we have calculated the branching ratios of the most important
channels.
It is possible to parameterize the effects of new physics beyond the
standard model by a series of effective operators $\mathcal{O}$
constructed with the standard model and the Majorana neutrino fields
and respecting the Standard Model $SU(2)_L \otimes U(1)_Y$ gauge
symmetry \cite{wudka}. These effective operators represent the
low-energy limit of an unknown theory. Their effects are suppressed
by inverse powers of the new physics scale $\Lambda$ for which we
take the value $\Lambda = 1 \; TeV$. Here we consider the effect of
dimension $6$ operators which are the dominant.

The total lagrangian is organized as follows:
\begin{eqnarray}
\mathcal{L}=\mathcal{L}_{SM}+\sum_{\mathcal{J},i}\frac{\alpha^{(i)}_{\mathcal{J}}}{\Lambda^{2}}
\mathcal{O}_\mathcal{J}^{i}
\end{eqnarray}
where $\mathcal{J}$ and $i$ labels the operators and families
respectively. For the considered operators we follow Ref
\cite{Aguila1} starting with a rather general effective lagrangian
density for the interaction of a Majorana neutrino $N$ with leptons
and quarks. All the operators which we list here are of the
dimension $6$ and could be generated at tree level in the unknown
fundamental high energy theory:
\begin{eqnarray}
\mathcal{O}^i_{LN\phi}=(\phi^{\dag}\phi)(\bar L_i N \tilde{\phi}),
\;\; \mathcal{O}^i_{NN\phi}=i(\phi^{\dag}D_{\mu}\phi)(\bar N
\gamma^{\mu} N), \;\; \mathcal{O}^i_{Ne\phi}=i(\phi^T \epsilon
D_{\mu} \phi)(\bar N \gamma^{\mu} e_i)
\end{eqnarray}
and for the baryon-number conserving 4-fermion contact terms we
have:
\begin{eqnarray}
\mathcal{O}^i_{duNe}=(\bar d_i \gamma^{\mu} u_i)(\bar N \gamma_{\mu}
e_i) &,& \;\; \mathcal{O}^i_{fNN}=(\bar f_i \gamma^{\mu} f_i)(\bar N
\gamma_{\mu}
N), \\
\mathcal{O}^i_{LNLe}=(\bar L_i N)\epsilon (\bar L_i e_i)&,& \;\;
\mathcal{O}^i_{LNQd}=(\bar L_i N) \epsilon (\bar Q_i d_i), \\
\mathcal{O}^i_{QuNL}=(\bar Q_i u_i)(\bar N L_i)&,& \;\;
\mathcal{O}^i_{QNLd}=(\bar Q_i N)\epsilon (\bar L_i d_i),\\
\mathcal{O}^i_{LN}=|\bar N L_i|^2&&
\end{eqnarray}
where $e_i$, $u_i$, $d_i$ and $L_i$, $Q_i$ denote the right handed
$SU(2)$ singlet and the left-handed $SU(2)$ doublets, respectively.

The operators listed above contribute to the effective lagrangian
\begin{eqnarray}\label{leff}
\mathcal{L}^N_{eff}&=&\frac{1}{\Lambda^2}\left\{\frac{v^2}{2}\alpha^{(i)}_{\phi}
\bar \nu_{L,i} N_R \Phi - \frac{M_Z^2 v}{2} \alpha_z \; Z^{\mu} \;
\bar N_R \gamma^{\mu} N_R - \frac{M_W^2 v}{\sqrt{2}} \alpha^{(i)}_W
\; W^{\dag\; \mu} \; \bar N_R \gamma_{\mu} e_{R,i} + \right.
\nonumber
\\ && \left. \alpha^{(i)}_{V_0} \bar d_{R,i} \gamma^{\mu} u_{R,i} \bar N_R \gamma_{\mu}
e_{R,i} + \alpha^{(i)}_{V_1} \bar e_{R,i} \gamma^{\mu} e_{R,i} \bar
N_R \gamma_{\mu} N_R + \alpha^{(i)}_{V_2} \bar L_i \gamma^{\mu} L_i
\bar N_R \gamma_{\mu} N_R + \right. \nonumber
\\ && \left. \alpha^{(i)}_{V_3} \bar u_{R,i} \gamma^{\mu}
u_{R,i} \bar N_R \gamma_{\mu} N_R + \alpha^{(i)}_{V_4} \bar d_{R,i}
\gamma^{\mu} d_{R,i} \bar N_R \gamma_{\mu} N_R + \alpha^{(i)}_{V_5}
\bar Q_i \gamma^{\mu} Q_i \bar N_R \gamma_{\mu} N_R + \right.
\nonumber
\\ && \left.
\alpha^{(i)}_{S_0}(\bar \nu_{L,i}N_R \bar e_{L,i}e_{R,i}-\bar
e_{L,i}N_R \bar \nu_{L,i}e_{R,i}) + \alpha^{(i)}_{S_1}(\bar
u_{L,i}u_{R,i}\bar N \nu_{L,i}+\bar d_{L,i}u_{R,i} \bar N e_{L,i})
 + \right. \nonumber
\\ && \left.
\alpha^{(i)}_{S_2} (\bar \nu_{L,i}N_R \bar d_{L,i}d_{R,i}-\bar
e_{L,i}N_R \bar u_{L,i}d_{R,i}) + \alpha^{(i)}_{S_3}(\bar u_{L,i}N_R
\bar e_{L,i}d_{R,i}-\bar d_{L,i}N_R \bar \nu_{L,i}d_{R,i}) + \right.
\nonumber
\\ && \left.  \alpha^{(i)}_{S_4} (\bar N_R \nu_{L,i} \bar \nu_{L,i} N_R+\bar
N_R e_{L,i} \bar e_{L,i} N_R)  + h.c. \right\}
\end{eqnarray}
where the sum over $i$ is understood  and the constants
$\alpha^{(i)}_{\mathcal{J}}$ are associated to specific operators
\begin{eqnarray}
\alpha_Z&=&\alpha_{NN\phi},\;
\alpha^{(i)}_{\Phi}=\alpha^{(i)}_{LN\Phi},\;
\alpha^{(i)}_W=\alpha^{(i)}_{Ne\Phi},\;
\alpha^{(i)}_{V_0}=\alpha^{(i)}_{duNe},\;\;
\alpha^{(i)}_{V_1}=\alpha^{(i)}_{eNN},\;\nonumber \\
\alpha^{(i)}_{V_2}&=&\alpha^{(i)}_{LNN},\;\alpha^{(i)}_{V_3}=\alpha^{(i)}_{uNN},\;
\alpha^{(i)}_{V_4}=\alpha^{(i)}_{dNN},\;\alpha^{(i)}_{V_5}=\alpha^{(i)}_{QNN},\;
\alpha^{(i)}_{S_0}=\alpha^{(i)}_{LNe},\;\nonumber \\
\alpha^{(i)}_{S_1}&=&\alpha^{(i)}_{QuNL},\;
\alpha^{(i)}_{S_2}=\alpha^{(i)}_{LNQd},\;\;
\alpha^{(i)}_{S_3}=\alpha^{(i)}_{QNLd},\;
\alpha^{(i)}_{S_4}=\alpha^{(i)}_{LN}
\end{eqnarray}
We calculate the cross-section for the production of the Majorana
neutrino according to the process shown in Fig.\ref{fig:eeNN}, valid
for the kinematic range $m_N < \sqrt{s}/2$
\begin{eqnarray}
{\large \sigma}^{NN}=\frac{s \beta}{64 \pi \Lambda^4}(\mathcal{C}^-
+ \mathcal{C}^+)(1+\frac{\beta^2}{3})
\end{eqnarray}
where
\begin{eqnarray}
\mathcal{C}^- &=& \left(\frac{M_Z^2 \alpha_Z}{(s-M_Z^2)} C_L +
\alpha^{(1)}_{V_2} + \alpha^{(1)}_{S_4}/2\right)^2 \;\;\;\;
 \mathcal{C}^+=\left(\frac{M_Z^2 \alpha_Z}{(s-M_Z^2)} C_R +
\alpha^{(1)}_{V_1} \right)^2 \nonumber \\
\beta &=& \sqrt{1-\frac{4 m_N^2}{s}}
\end{eqnarray}
and $C_L = -1/2+x_W$, $C_R=x_W$, with $x_W=\sin^2\theta_W$.
\begin{figure*}[htbp]
\centering
\includegraphics[angle=270,totalheight=6cm,bb= 70 180 550 680]{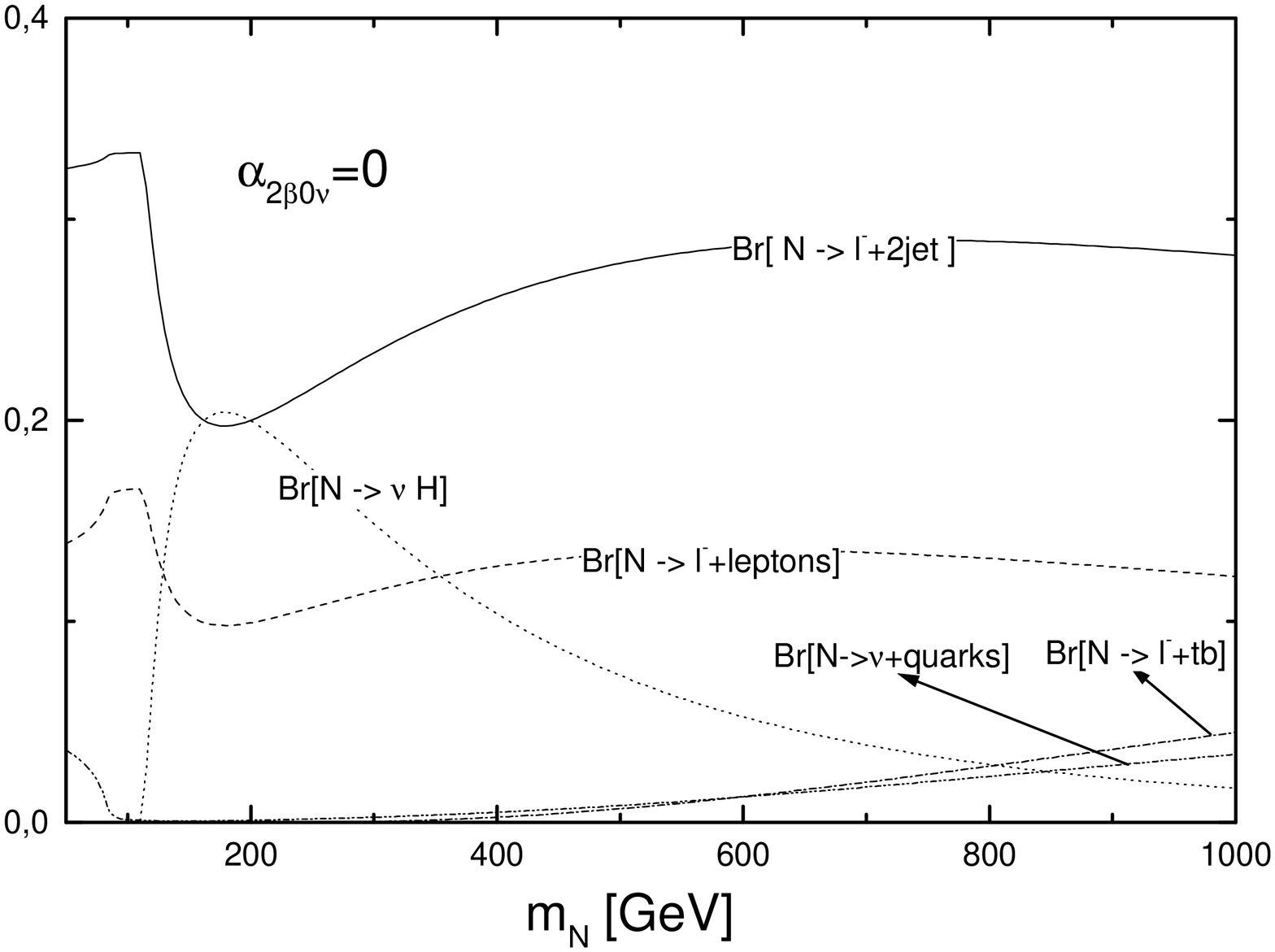}
\hspace{2cm}
\includegraphics[angle=270,totalheight=6cm,bb= 70 180 550 680]{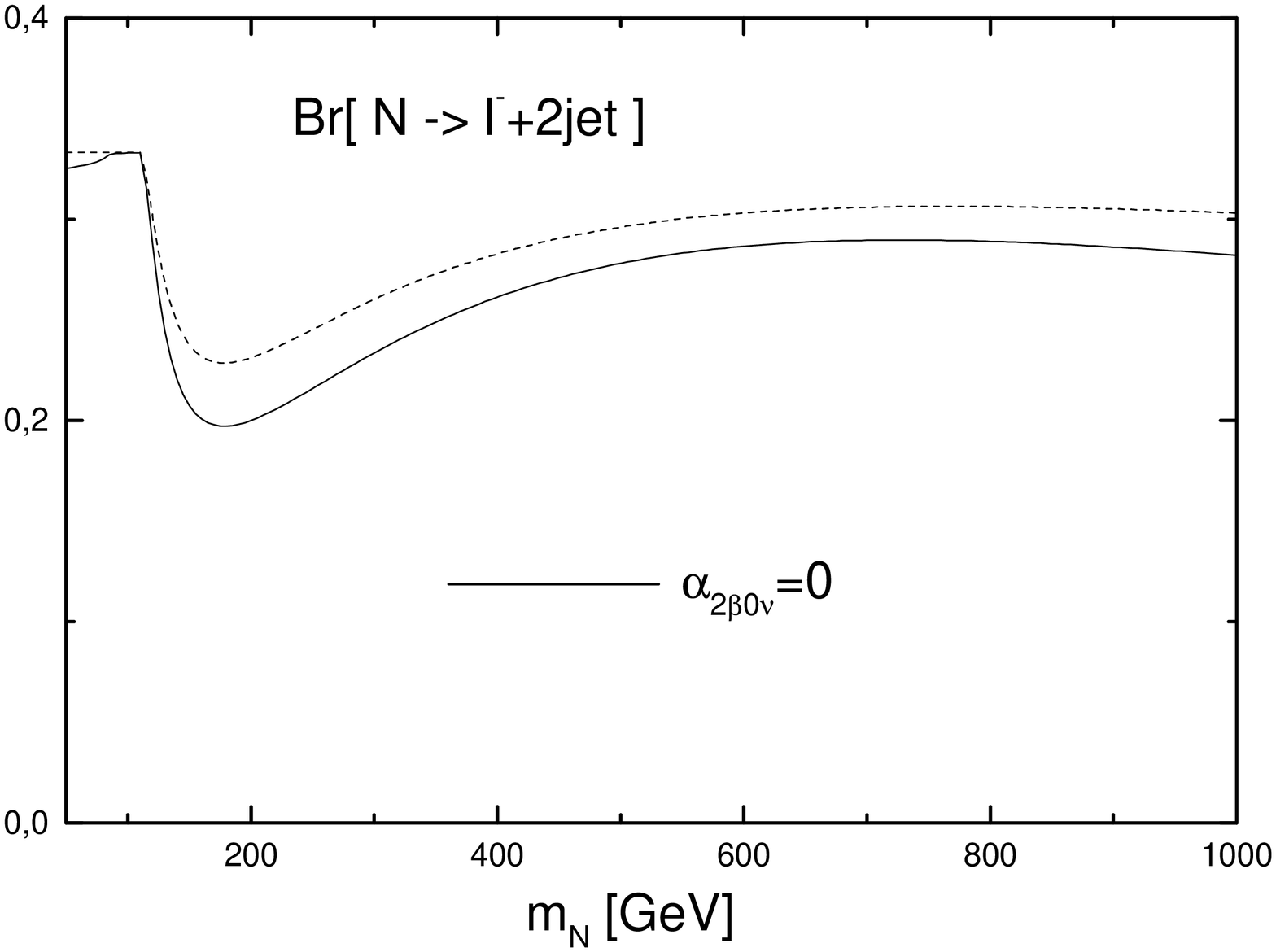}
\vspace{0.5cm} \caption{\label{fig:br_majo}  In the left panel the
Branching ratios for the Majorana neutrino decay with
$\alpha_{0\nu\beta\beta}=0$. In the right panel there is a
comparison with the same-coupling case for the $N \rightarrow l^- +
jets$ decay.}
\end{figure*}
We also study the single Majorana neutrino production. The
corresponding diagrams are shown in Fig.\ref{fig:eeN} and the result
for the square amplitud is
\begin{eqnarray}
|\mathcal{\bar M}|^2 &=&
\frac{g^2}{8\Lambda^2}\frac{1}{(q^2)^2}\left\{8\alpha_w^2
\frac{\Pi_w^2}{M_w^2} s [2 (k\cdot p) (2 (k \cdot q) (l \cdot q) -
(k \cdot l) q^2)+M_w^2(2 (l \cdot q) (p \cdot q) - (l \cdot p) q^2)]
\right. \nonumber \\ &+& \left. 4(\alpha^{(1)}_{S_0})^2 \frac{(l
\cdot p_2)}{M_w^2}[2 (k \cdot p) ( 2 (k \cdot q) (p_1 \cdot q) - (k
\cdot p_1) q^2) \right. \nonumber \\ &+& \left. M_w^2 (2 (p \cdot q)
(p_1 \cdot q) - (p \cdot p_1) q^2)] \right\}
\end{eqnarray}
where $q=p+k$ and $p_1$, $p_2$, $l$, $p$ and $k$ are the 4-momenta
of the electron, the positron, the Majorana neutrino $N$, the
charged lepton and the $W$ boson respectively. The $W$ propagator is
$\Pi_W=m_W^2/[(p_1-l)^2-m_W^2]$.

\begin{figure*}[htbp]
\begin{center}
\includegraphics[angle=270,totalheight=6cm,bb= 70 180 550 680]{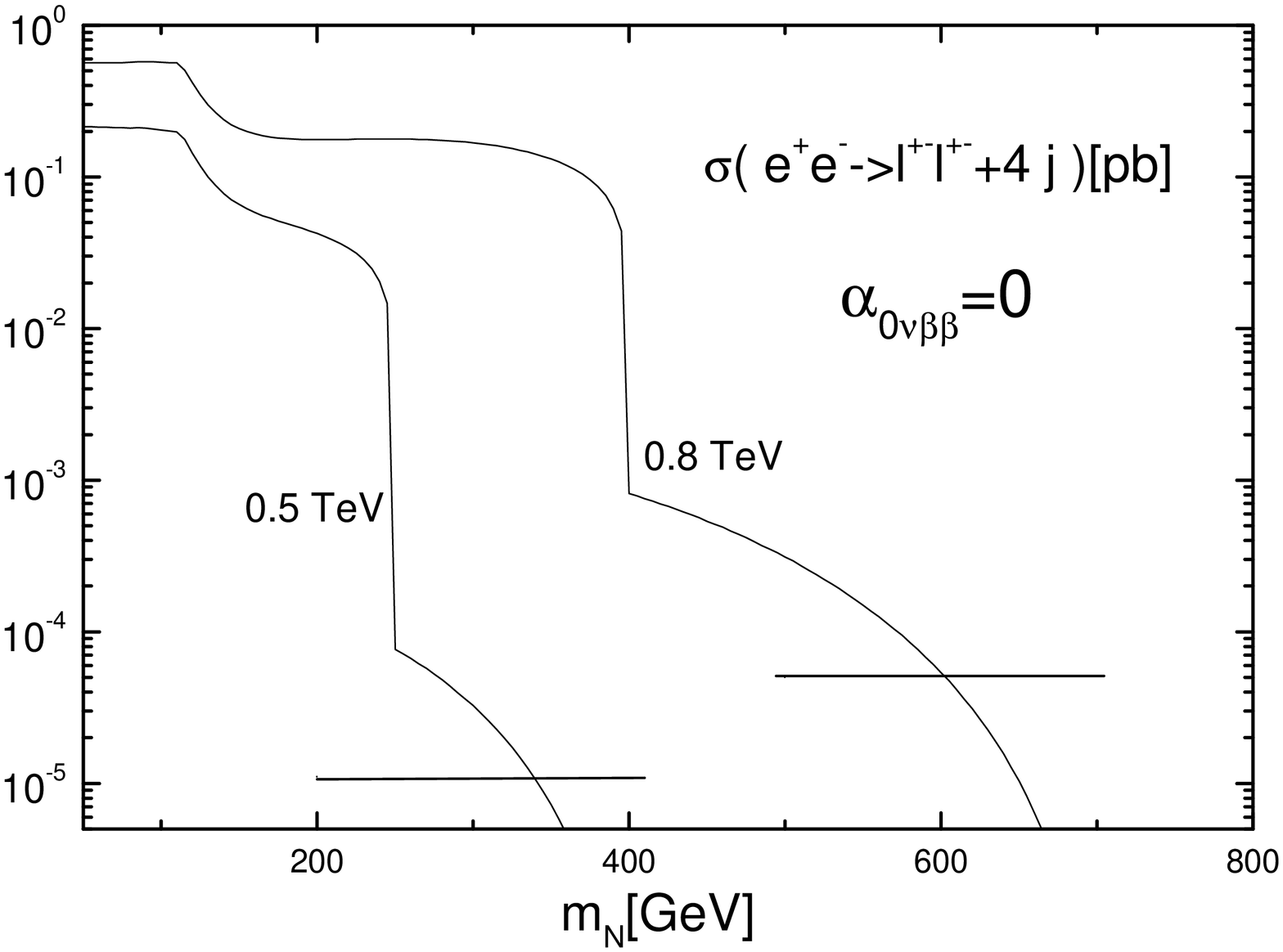}
\hspace{2cm}
\includegraphics[angle=270,totalheight=6cm,bb= 70 180 550 680]{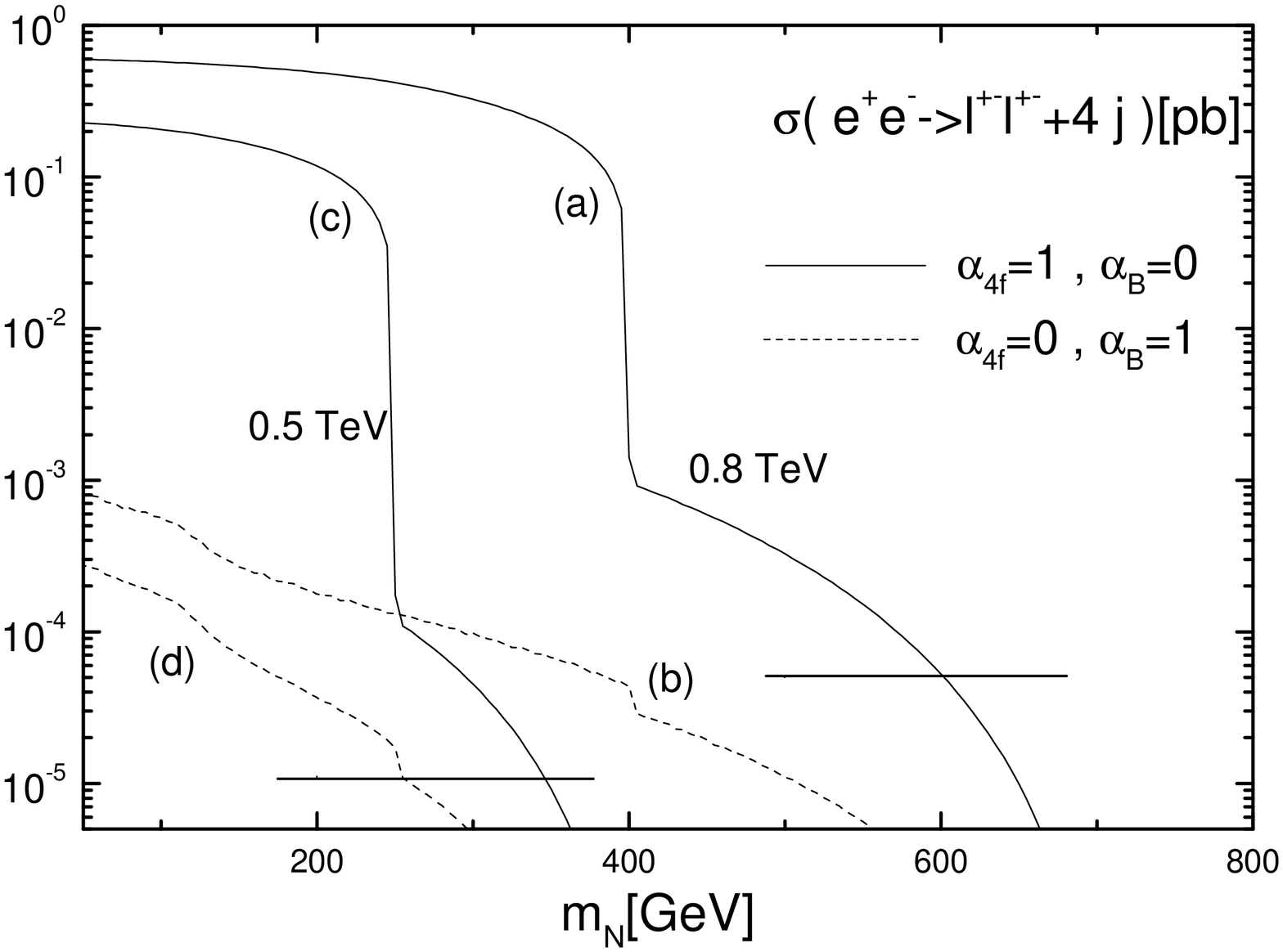}
\vspace{0.5cm}
 \caption{\label{fig:crosssection} The cross-section for the process $e^+e^- \rightarrow l^{\pm}l^{\pm}+jets$. The left panel correspond
 to the zero contribution of the operators related with $0\nu_{\beta\beta}$ ($\alpha_{0\nu_{\beta \beta}}=0$). In the right panel
 the curves labeled (a), (b), (c) and (d) correspond
 to energies of center of mass $\sqrt{s}=0.5TeV$,
 $\sqrt{s}=0.8TeV$ and different values of the constant $\alpha^{(i)}_{\mathcal{J}}$.
 In both panels the horizontal solid line represent the values of the SM background.}
\end{center}
\end{figure*}

The cross section is obtained by integrating the phase space in the
usual way, using the numerical routine RAMBO \cite{rambo}.

The total cross section is the combination of the two processes
mentioned above in the approximated expression:
\begin{eqnarray}
\label{sigtot}
 \sigma^{(e^+e^- \rightarrow l^{\pm} l^{\pm}  jets)}
&=& 2 \left( \sum_{i,j} \sigma^{(e^+e^- \rightarrow NN)}
Br(N\rightarrow l_i^+ +
 jets) Br(N\rightarrow l_j^+ +  jets) \Theta(\sqrt{s}/2-m_N)
\right. \nonumber \\  &+& \left. \sum_i \sigma^{(e^+e^- \rightarrow
N e^+ W)} Br(N \rightarrow l_i^+  jets) Br(W \rightarrow  jets)
\Theta(\sqrt{s}-m_N) \right)
\end{eqnarray}
The factor two in front of Eq.(\ref{sigtot}) takes into account the
possible charges of the final leptons.

As we shall see later, some of the considered operators contribute
to the neutrinoless double beta decay ($0\nu_{\beta\beta}$-decay)
and may be strongly constrained. In these conditions we analyze the
case where the  operators contributing to the
$0\nu_{\beta\beta}$-decay vanishes whist the rest are non-zero and
contribute with similar strength. For completion we also analyze the
case where all of the operators contribute with similar strength.
The branching ratios shown in the left panel of
Fig.(\ref{fig:br_majo}) (the expressions are collected in the
Appendix) correspond to the case with non contribution of the
$0\nu_{\beta\beta}$-decay related operators
($\alpha_{0\nu\beta\beta}=0$). In the right panel of the same figure
we show for comparison the branching ratio for $N \rightarrow l^+ +
jets$ with $\alpha_{0\nu\beta\beta}=0$ and
$\alpha_{0\nu\beta\beta}\neq 0$. As we can see there are not
significant differences.

In Fig.\ref{fig:crosssection} we show the results for the cross
section combining the processes shown in Figs.\ref{fig:eeNN} and
\ref{fig:eeN} with the $W$-boson decaying into hadrons and the
Majorana neutrino $N$ decaying according to the Branching Ratios
shown in the Appendix. We show the result as a function of the
Majorana neutrino mass $m_N$ and center of mass energies of
$\sqrt{s}=0.5 \; TeV$ and $\sqrt{s}=0.8 \; TeV$. We have considered
$\sqrt{s} < \Lambda$ in order to ensure the validity of the
effective lagrangian approach. The cross section is calculated for
different values of the constants $\alpha^{(i)}_{\mathcal J}$. In
the left panel of the Fig.\ref{fig:crosssection} we have shown the
cross section for the case in which the constants related with the
operators contributing to $0\nu_{\beta\beta}$-decay are considered
zero. These operators are $\mathcal{O}^1_{N e \phi}$,
$\mathcal{O}^1_{d u N e}$, $\mathcal{O}^1_{Q u N L}$,
$\mathcal{O}^1_{L N Q d}$ and $\mathcal{O}^1_{Q N L d}$, as will be
discussed in section \ref{sec:2beta}. In the right panel we plot the
cross section where the operators that contribute are the 4-fermion
operators (solid line) or the operators involving bosons (dashed
line). In both panels, the non-zero coupling constants take the
value one. As we can see, the 4-fermion contribution is the most
important. In both panels we show with a horizontal solid line the
value of the SM background as we will be explained later in the
text.

The final leptons can be either of $e^{\pm}$, $\mu^{\pm}$ or
$\tau^{\pm}$ since this is allowed by the interaction lagrangian
(Eq.\ref{leff}). All of these possible final states are clear
signals for intermediary Majorana neutrinos, thus we sum the cross
section over the flavors of the final leptons.  The partial width of
$N$ was determined at tree level considering the dominant decay
modes $N\rightarrow l + 2 jets$, $N\rightarrow l + tb$,
$N\rightarrow l + leptons$, $N\rightarrow \nu + H$ and $N\rightarrow
\nu + quarks$ coming from the higgs, the charged $W$-boson and the
4-fermion effective interactions. We present in the Appendix the
differential partial-width for its dominant decay channels, where
the contributing effective operators are identified by the indicated
labels in the couplings.

\section{The Standard Model Background and the Discovery Region}

The considered signal is strictly forbidden in the Standard Model.
The SM background, which can be confused with the studied reaction,
will always involve additional light neutrinos. The dominant SM
process arises from the resonant production of four $W^{\pm}$
bosons: $e^+e^- \rightarrow W^+W^+W^-W^-$, the decay of two
$W^\prime$s into leptons $W^{\pm} \rightarrow l^{\pm} + \nu (\bar
\nu)$, and the other two into jets, $W \rightarrow jets$. We
calculated the cross section for these processes using the package
COMPHEP \cite{comphep1,comphep2,comphep3} and we multiplied it by
the corresponding branching ratios (BR$[W\rightarrow l \nu])^2
\simeq 0.011$ and the (Br$[W \rightarrow 2 jets])^2\simeq 0.46$ and
by the factor $18$ to take into account the different combinations
of the same sign charged final leptons: $l^{\pm}l^{\pm}$ with
$l=e,\mu,\tau$. The calculated values are $5.0 \; 10^{-5}$ pb and
$1.1 \; 10^{-5}$ pb for $\sqrt{s}=0.8$ TeV and $0.5$ TeV
respectively. In Table \ref{tabla1} we compare the values of the
signal, for different values of the Majorana neutrino mass, with the
SM background. In Fig.\ref{fig:crosssection} we show along, with the
signal cross-section, the corresponding background levels as
horizontal lines for $\sqrt{s}=0.5$TeV and $\sqrt{s}=0.8$TeV.

\begin{figure*}[htbp]
\begin{center}
\includegraphics[totalheight=5cm]{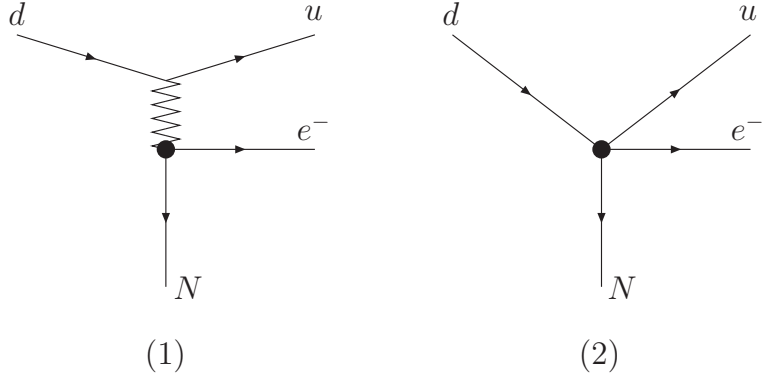}
\caption{\label{fig:2betaope} Contribution to $\mathcal H$ in
Eq.(\ref{heff}). In the diagram (1) the solid dot represent the
operator ${\mathcal O}^1_{Ne\phi}$ and in the diagram (2) the
4-fermion operators ${\mathcal O}^1_{duNe}$, ${\mathcal
O}^1_{QuNL}$,
 ${\mathcal O}^1_{LNQd}$ and ${\mathcal O}^1_{QNLd}$}.
\end{center}
\end{figure*}

\begin{figure*}[htbp]
\begin{center}
\includegraphics[angle=270,totalheight=6cm,bb= 70 180 550 680]{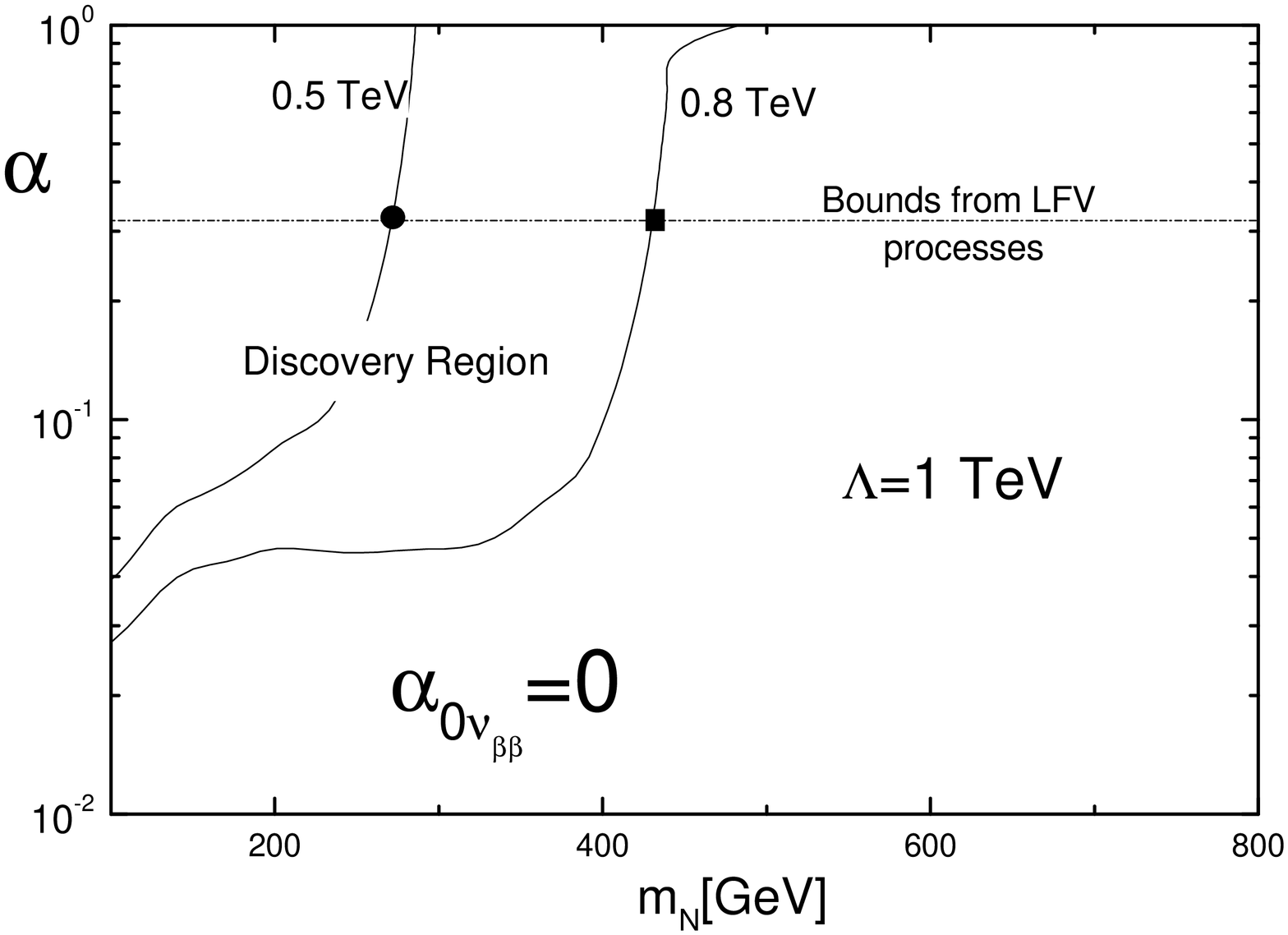}
\hspace{2cm}
\includegraphics[angle=270,totalheight=6cm,bb= 70 180 550 680]{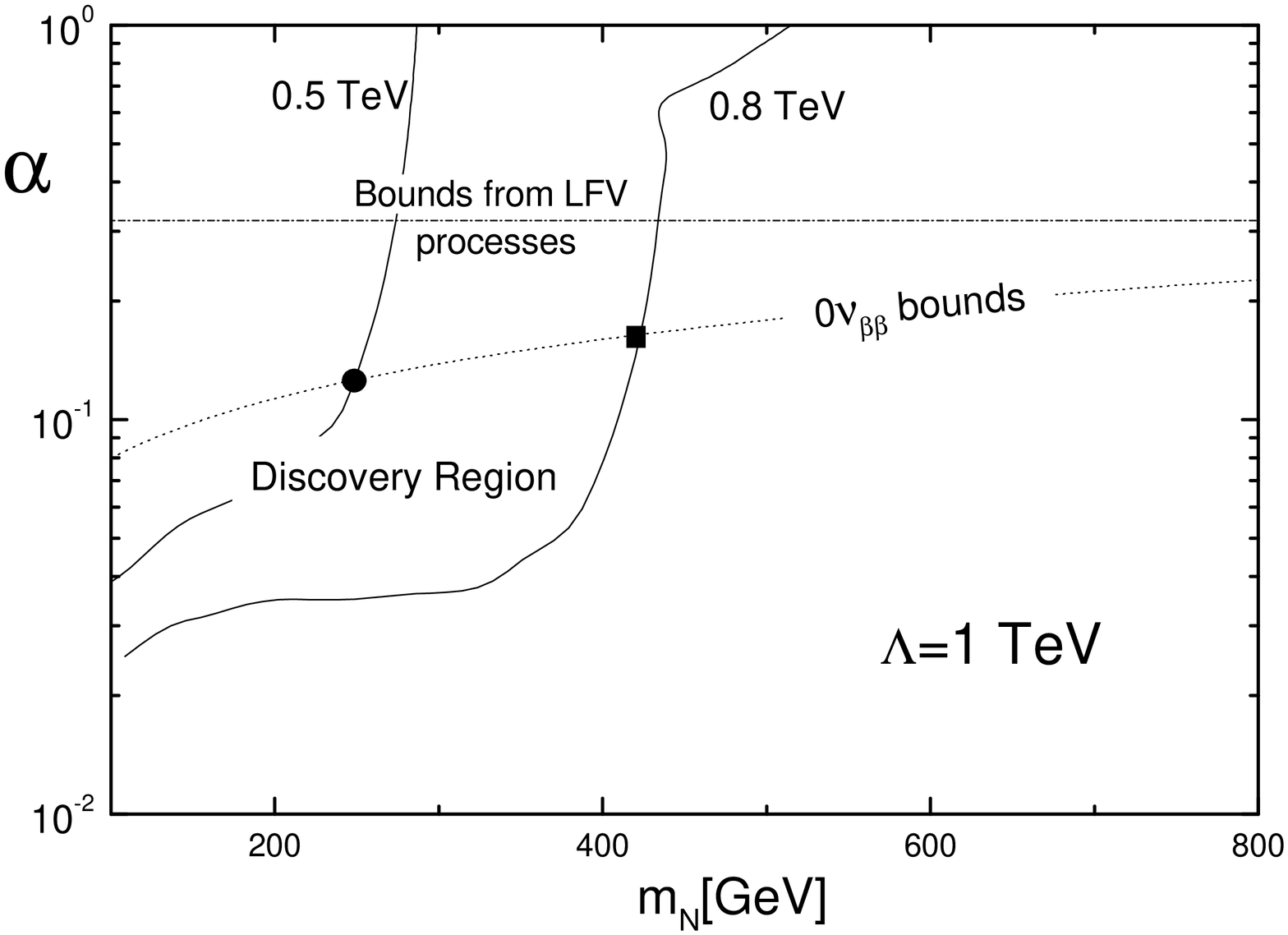}
\vspace{0.5cm}
 \caption{\label{fig:discovery} Discovery region above the solid
curve at 5 $\sigma$ (this work) and below the dot-dashed line for
LVF bound (Left panel) or the dotted curve for
$0\nu_{\beta\beta}$-decay bound (Right panel).}
\end{center}
\end{figure*}

In order to investigate the capability of the studied process to
discover effects of Majorana neutrinos, we study the region
(discovery region) where the signal can be separated from the
background with a statistical significance of $5 \sigma$. It is done
by defining the quantity $\mathcal{S}$
\begin{eqnarray}
\mathcal{S}=\frac{L[\sigma(\alpha,M_N)-\sigma_{B}]}{\sqrt{L[\sigma(\alpha,M_N)+\sigma_{B}]}}
\end{eqnarray}
where $L$ is the luminosity and the numerator represents the
discrepancy between the signal and the SM background ($\sigma_B$).
In Fig.\ref{fig:discovery} we show the discovery region (above the
solid curves) where $\mathcal{S} \geq 5$ (5 $\sigma$ statistical
significance) for a luminosity $L=100 fb^{-1}$.

For completion we have also considered, although in an approximated
way, the bounds on the operators which come from the
$0\nu_{\beta\beta}$-decay and from LEP and low energy data. The
former will be considered in the next section and the latter in the
following.

The heavy Majorana neutrino couples to the three flavors families
with couplings dependent on the scale $\Lambda$ and the constants
$\alpha^{(i)}_{\mathcal{J}}$.  It is possible to relate this
coupling with the mixing between light and heavy neutrinos
($U_{eN}$, $U_{\mu N}$, $U_{\tau N}$) for which the experimental
bounds, obtained from LEP and low energy data, have been put in
\cite{Aguila2,pilaftis,langaker,roulet}. This relation was found in
\cite{Aguila1} comparing the operator $\mathcal{O}^i_{Ne\phi}$ with
the strength of the V-A interaction (Eq.(\ref{lw})). It is $U_{l_i
N} \simeq (\alpha^{(i)}_{W}/2)( v^2/\Lambda^2)$ where $v$
corresponds to the vacuum expectation value: $v=250 GeV$.  In order
to keep the analysis as simple as possible we consider that the same
bound applies for all the couplings $\alpha^{(i)}_{\mathcal{J}}$,
generically $\alpha$.

In our case, for one heavy Majorana neutrino, and following
\cite{Aguila2}, we have:
\begin{eqnarray}
\Omega_{l l'} = U_{l N} U_{l' N}
\end{eqnarray}
where the allowed values for the parameter are \cite{bergmann}:
\begin{eqnarray}
\Omega_{e e}\leq 0.0054, \;\; \Omega_{\mu \mu}\leq 0.0096, \;\;
\Omega_{\tau \tau} \leq 0.016
\end{eqnarray}
For the Lepton-Flavour-Violating process (LFV), e.g. $\mu
\rightarrow e \gamma$, $\mu \rightarrow e e e$ and $\tau \rightarrow
e e e$, which are induced by the quantum effect of the heavy
neutrinos, we have \cite{tommasini}:
\begin{eqnarray}
|\Omega_{e \mu}| \leq 0.0001, \;\; |\Omega_{e \tau}| \leq 0.02, \;\;
|\Omega_{\mu \tau}| \leq 0.02
\end{eqnarray}
These bounds can be translated to the constants $\alpha$
considering, in a simplified way, that all the operators satisfy the
same constraint
\begin{eqnarray}
\Omega_{e \mu}=U_{e N} U_{\mu N} = \left(\frac{\alpha}2
\frac{v^2}{\Lambda^2} \right)^2 < 0.0001
\end{eqnarray}
For $\Lambda=1$ TeV we have:
\begin{eqnarray}
\alpha \leq 0.32
\end{eqnarray}
This value is shown in both panels of Fig.\ref{fig:discovery} with a
horizontal dot-dashed line.

\begin{center}
\begin{table}
\begin{tabular}{|c|c|c|c|}
\hline
$\sqrt{s}(GeV)$ & $ \sigma_{SM} (pb) $ & $ M_N (GeV)$ &  $ \sigma (pb)$  \\
\hline \hline
        &            & 100 & $ 0.2 $ \\
 500    & $1.1 \; 10^{-5}$ & 200 & $ 0.043 $ \\
        &            & 300 & $ 3.3\;10^{-5} $ \\
\hline \hline
        &            & 200 & $ 0.5 $ \\
 800    &$ 5.0 \; 10^{-5}$ & 300 & $ 0.33 $ \\
        &            & 500 & $ 3.2\;10^{-4} $ \\
 \hline
\end{tabular}
\caption{ Comparison between the signal and background.}
\label{tabla1}
\end{table}

\end{center}

\section{Neutrinoless double beta decay bounds}\label{sec:2beta}

In order to take into account the bounds imposed by the
$0\nu_{\beta\beta}$-decay experiment on some of the coupling
constants $\alpha^{(i)}_{\mathcal J}$, we consider, in a general
way, the following effective interaction Hamiltonian:
\begin{equation}
\label{heff}
\mathcal{H}=G_{eff} \; \bar u \Gamma d \;\; \bar e
\Gamma N + h.c.
\end{equation}
where $\Gamma$ represents a general Lorentz-Dirac structure.
Following the developments presented in \cite{mohapatra,rodejohann}
and using the most stringent limits on the lifetime for neutrinoless
double beta decay $\tau_{{0\nu}_{\beta\beta}} \geq 1.9 \times
10^{25}$ yr obtained by the Heidelberg-Moscow Collaboration
\cite{HMColla}, we have obtained the following bounds for $G_{eff}$
\begin{equation}
G_{eff} \leq 8.0 \times 10^{-8} \left( \frac{m_N}{100 GeV}
\right)^{1/2} GeV^{-2}
\end{equation}
The lowest order contribution to $0\nu_{\beta \beta}$-decay from the
considered effective operators comes from those that involve the $W$
field and the 4-fermion operators with quarks $u$, $d$, the lepton
$e$ and the Majorana neutrino $N$:
\begin{eqnarray}
\label{2beope}
 \mathcal O^1_{N e \phi} \;,\; \mathcal O^1_{duNe} \;,\;
\mathcal O^1_{QuNL} \;,\; \mathcal O^1_{LNQd} \;,\; \mathcal
O^1_{QNLd}
\end{eqnarray}
The contribution of these operators to the effective Hamiltonian in
eq.(\ref{heff}) is shown in Fig.\ref{fig:2betaope}.

For the coupling constant associated with each operator we use the
generic name $\alpha_{0\nu_{}\beta\beta}$, that is to say
\begin{eqnarray}
\alpha_{0\nu\beta\beta}=\alpha^{(1)}_{Ne\phi}=\alpha^{(1)}_{duNe}=\alpha^{(1)}_{QuNL}=\alpha^{(1)}_{LNQd}=\alpha^{(1)}_{QNLd}
\end{eqnarray}

In order to estimate the bounds on the different $\alpha_{\mathcal
J}^{(i)}$ we consider two different situations. First, we suppose
that the contribution of all the operators involved in the
$0\nu_{\beta\beta}$-decay adds constructively. In this case we
expect strong limits on the couplings, then we may assume them to be
negligible. Thus, we consider that all of the constants associated
with the operators in Eq.(\ref{2beope}) vanish
($\alpha_{0\nu\beta\beta}=0$) and that the other constants, which
are not bounded by neutrinoless double beta decay, are non-zero and
have similar magnitude. This situation is shown in the left panel of
the Fig.(\ref{fig:discovery}).

Second, we consider the individual contributions of each operator as
acting alone. In this case it is obvious to relate the coupling with
the $G_{eff}$ in Eq.(\ref{heff})
\begin{equation}
G_{eff}=\frac{\alpha_{0\nu\beta\beta}}{\Lambda^2}
\end{equation}
Thus we can translate the limit which came from  $G_{eff}$ to
$\alpha_{0\nu\beta\beta}$. For, $\Lambda=1 TeV$, it is
\begin{equation}
\alpha_{0\nu\beta\beta} \leq 8.0 \times 10^{-2}
\left(\frac{m_N}{100GeV}\right)^{1/2}
\end{equation}
Taking a conservative point of view, in the right panel of
Fig.\ref{fig:discovery}, we present this bound considering that it
is the same for all the constants $\alpha^{(i)}_{\mathcal{J}}$
(generically $\alpha$) and show it with the dotted curve. The solid
curve, which is the contribution of this work, represent the lower
limit for the discovery region. In the right panel, it was
calculated considering that all the constants
$\alpha^{(i)}_{\mathcal J}$ have similar magnitude. On the other
hand, in the left panel it was calculated considering
$\alpha_{0\nu\beta\beta}=0$. We also show the bound from Lepton
Flavors Violating process with the dot-dashed line in the same
figure.

 Summarizing, we calculated the cross-section for the process
$e^+ e^- \rightarrow l_j^\mp l_k^\mp +  \; jets $ where $l_1$, $l_2$
and $l_3$ are light leptons $(e,\mu,\tau)$ respectively. We show the
total unpolarized cross-section using the calculated Branching
ratios for different values of $m_N$ and the coupling
$\alpha^{(i)}_{\mathcal J}$. We showed the discovery regions at
5$\sigma$ statistical significance combining with the
$0\nu_{\beta\beta}$ and the LFV bounds. We found that it will be
possible to discover Majorana neutrinos with masses lower than $250
\; GeV$ and $400 \; GeV $ at $e^+e^-$ colliders with center of mass
energy of $0.5 \; TeV$ and $0.8 \; TeV$ respectively.

{\bf Acknowledgements}

We thank CONICET (Argentina) and Universidad Nacional de Mar del
Plata (Argentina) for their financial supports.

\section{Appendix}

We present here the partial decay widths of a heavy Majorana
neutrino N for its dominant decay channels. They were calculated
using the effective interactions shown above in the text.

\begin{eqnarray}
\frac{d\Gamma}{dx}^{(N\rightarrow l^{+} \bar u d)}&=&\frac{m_N }{256
\pi^3}\left(\frac{m_N}{\Lambda}\right)^4  x^2
 \left\{ \left[\frac32 \sum_{i=1,2}(\alpha_{s_1,i}^2+\alpha_{s_2,i}^2-\alpha_{s_2,i}\alpha_{s_3,i})
   (1-x)  \right. \right. \nonumber \\  &+&  \left. \left.
\frac14
\sum_{i=1,2}(\alpha_{s_3,i}^2+4\alpha_{V_0,i}^2)(3-2x)\right]
\right.  \nonumber \\ &+& \left.
 \left( \sum_{i=1,3}
\alpha_{W,i}^2\right) \left(\frac2{w+(1-(1-x)z)^2}\right)(3-2x)
\right\} \;\; \mbox{with} \;\; 0<x<1 \nonumber \\
\mbox{\vspace{1cm}} \nonumber \\ \mbox{\vspace{1cm}}
\frac{d\Gamma}{dx}^{(N\rightarrow l^{+} \bar t b)}&=&\frac{m_N }{256
\pi^3}\left(\frac{m_N}{\Lambda}\right)^4 \frac{(1-x-y)^2
x^2}{(1-x)^3}
 \left\{ \left[\frac32(\alpha_{s_1,3}^2+\alpha_{s_2,3}^2-\alpha_{s_2,3}\alpha_{s_3,3})
   (1-x)^2  \right. \right. \nonumber \\  &+&  \left. \left.
\frac14(\alpha_{s_3,3}^2+4\alpha_{V_0,3}^2)((3-2x)(1-x)+y(3-x))\right]
\right.  \nonumber \\ &+& \left.
\left(\sum_{i=1,3}\alpha_{W,i}^2\right)\frac{(3-2x)(1-x)+y(3-x)
}{w+(1-(1-x)z)^2} \right\} \;\; \mbox{with} \;\; 0<x<1-y \nonumber \\
\mbox{\vspace{1cm}} \nonumber \\ \mbox{\vspace{1cm}}
\frac{d\Gamma}{dx}^{(N \rightarrow \nu d d)}&=&\frac{m_N}{256\pi^3}
\left( \frac{m_N}{\Lambda}\right)^4
\frac{x^2}{4}\left[\sum_{i=1,3}\left(\alpha_{s_2,i}^2-\alpha_{s_2,i}\alpha_{s_3,i}\right)
6(1-x)  \right. \nonumber \\  &+&  \left.  \left( \sum_{i=1,3}
\alpha_{s_3,i}^2\right) (3-2x) \right] \;\; \mbox{with} \;\; 0<x<1
\nonumber \\ \mbox{\vspace{1cm}} \nonumber \\ \mbox{\vspace{1cm}}
\frac{d\Gamma}{dx}^{(N \rightarrow \nu u
u)}&=&\frac{m_N}{256\pi^3}\left( \frac{m_N}{\Lambda}\right)^4 \left(
\sum_{i=1,3} \alpha_{s_1,i}^2\right) \frac32 x^2(1-x)  \;\;
\mbox{with} \;\; 0<x<1 \nonumber \\ \mbox{\vspace{1cm}} \nonumber \\
\mbox{\vspace{1cm}} \nonumber
\end{eqnarray}

\begin{eqnarray}
\frac{d\Gamma}{dx}^{(N \rightarrow \nu t
t)}&=&\frac{m_N}{256\pi^3}\left( \frac{m_N}{\Lambda}\right)^4
\left(\alpha_{s_1,3}^2\right) \frac32 x^2 \sqrt{1-\frac{4y}{1-x}}
(1-x-2y) \Theta(1-x-4y) \nonumber \\ \mbox{\vspace{1cm}} \nonumber \\
\mbox{\vspace{1cm}} &&\mbox{with} \;\; 0<x<1-4 y \nonumber \\
\mbox{\vspace{1cm}} \nonumber \\ \mbox{\vspace{1cm}}
\frac{d\Gamma}{dx}^{(N\rightarrow l^{+} leptons)}&=&\frac{m_N }{256
\pi^3}\left(\frac{m_N}{\Lambda}\right)^4  \frac{x^2}{12}(3-2x)
 \left[ \left(\sum_{i=1,3}\alpha_{s_0,i}^2\right) \right. \nonumber \\  &+&  \left.
\left(\sum_{i=1,3}\alpha_{W,i}^2\right)\frac{12}{w+(1-(1-x)z)^2}
\right]\;\; \mbox{with} \;\; 0<x<1 \nonumber \\ \mbox{\vspace{1cm}}
\nonumber \\ \mbox{\vspace{1cm}} \frac{d\Gamma}{dx}^{(N \rightarrow
\nu H)}&=&\frac{m_N}{256 \pi^3} \left(\frac{m_N}{\Lambda}\right)^4
\left(
\sum_{i=1,3}\alpha_{\phi,i}^2\right)\left(\frac{v}{m_N}\right)^4
2\pi^2 (1-z_{\phi}) \nonumber \\ \nonumber \;\; &&\mbox{with} \;\;
0<x<1
\end{eqnarray}

where  $x= 2 p^0_{lepton}/m_N$, and $y=(m_t/m_N)^2$,
$z_{\phi}=(m_{\phi}/m_N)^2$, $z=(m_N/m_W)^2$, $w=(\Gamma_w/m_W)^2$.

\pagebreak

\end{document}